\documentclass[12pt,letterpaper]{article}
\usepackage{graphicx}

\begin{document}

\title{ {\bf The Emergence of Classical Dynamics in a Quantum World } }

\author{Tanmoy Bhattacharya, Salman Habib, and Kurt Jacobs}
\date{}
\maketitle
\thispagestyle{empty}

\vspace{-1.5cm}
\begin{center}
\hspace{0.7cm} {\em T-8, Theoretical Division, The University of
California,} 
\newline
{\em Los Alamos National Laboratory, Los Alamos, New Mexico 87545.}
\end{center}
\vspace{0cm}

Ever since the advent of quantum mechanics, it has been clear that the
atoms composing matter do not obey Newton's laws. Instead, their
behavior is described by the Schr\"odinger equation. Surprisingly
though, until recently, no clear explanation was given for why
everyday objects, which are merely collections of atoms, are observed
to obey Newton's laws. It would seem that, if quantum mechanics
explains all the properties of atoms accurately, they, too, should
obey quantum mechanics. This reasoning led some scientists to believe
in a distinct macroscopic, or ``big and complicated,'' world in which
quantum mechanics fails and classical mechanics takes over, although
there has never been experimental evidence for such a failure.  Even
those who insisted that Newtonian mechanics would somehow emerge from
the underlying quantum mechanics as the system became increasingly
macroscopic were hindered by the lack of adequate experimental and
theoretical tools. In the last decade, however, this
quantum-to-classical transition has become accessible to experimental
study and quantitative description, and the resulting insights are the
subject of this article~\footnote{The original version of this article 
was published in Los Alamos Science {\bf 27}, 110 (2002).}.

\vspace{-1.2cm}
\section*{\begin{center} The Quantum to Classical Transition \end{center}}
\vspace{-0.4cm}

We will illustrate the problems involved in describing the
quantum-to-classical transition by using the example of a baseball
moving through the air. Most often, we describe how the ball moves
through air, how it spins, or how it deforms. Regardless of which
degree of freedom we might consider -- whether it is the position of
the center of mass, angular orientation, or deviation from sphericity
-- in the final analysis, those variables are merely a combination of
the positions (or other properties) of the individual atoms. As all
the properties of each of these atoms, including position, are
described by quantum mechanics, how is it that the ball as a whole
obeys Newton's equations instead of some averaged form of the
Schr\"odinger equation?

Even more difficult to explain is how the chaotic behavior of
classical, nonlinear systems emerges from the behavior of quantum
systems. Classical, nonlinear dynamical systems exhibit extreme
sensitivity to initial conditions. This means that, if the initial
states (for example, particle positions and momenta) of two identical
copies of a system differ by some tiny amount, those differences
magnify with time at an exponential rate.  As a result, in a very
short time, the two systems follow very different evolutionary
paths. On the other hand, concepts such as precise position and
momentum do not make sense according to quantum mechanics: We can
describe the state of a system in terms of these variables only
probabilistically. The Schr\"odinger equation governing the evolution
of these probabilities typically makes the probability distributions
diffuse over time.  The final state of such systems is in general not
very sensitive to the initial conditions, and the systems do not
exhibit chaos in the classical sense.

The key to resolving these contradictions hinges on the following
observation: While macroscopic mechanical systems may be described by
single quantum degrees of freedom, those variables are subject to
observation and interaction with their environment, which are
continual influences. For example, a baseball's center-of-mass
coordinate is continually affected by the numerous properties of the
atoms composing the baseball, their thermal motion, random collisions
with air molecules, and even the light that reflects off it. The
process of observing the baseball's motion also involves interaction
with the environment: Light reflected off the baseball and captured by
the observer's eye creates a trace of the motion on the retina.

In the next section, we will show that, under conditions that refine
the intuitive concept of what is macroscopic, the motion of a quantum
system is basically indistinguishable from that of a classical system!
In effect, observing a quantum system provides information about it
and counteracts the inherent tendency of the probability distribution
to diffuse over time although observation creates an irreducible
disturbance.  In other words, as we observe the system continuously,
we know where it is and do not have to rely upon the progressively
imprecise theoretical predictions of where it could be. When one takes
into account this ``localization'' of the probability distribution
encoding our knowledge of the system, the equations governing the
expected measurement results (that is, the equations telling us what
we observe) become nonlinear in precisely the right way to recover an
approximate form of classical dynamics -- for example, Newton's laws
in the baseball example.

What happens when no one observes the system? Does the baseball
suddenly start behaving quantum mechanically if all observers close
their eyes? The answer is hidden in a simple fact: Any interaction
with a sufficiently complicated external world has the same effect as
a series of measurements whose results are not recorded. In other
words, the nature of the disturbance on the system due to the system's
interactions with the external world is identical to that of the
disturbance observed as an irreducible component of
measurement. Naturally, questions about the path of the baseball
cannot be verified if there are no observers, but other aspects of its
classical nature can, and do, survive.

\section*{Digression: A Historical Perspective}

The demands imposed by quantum mechanics on the disciplines of
epistemology and ontology have occupied the greatest minds. Unlike the
theory of relativity, the other great idea that shaped physical
notions at about the same time, quantum mechanics does far more than
modify Newton's equations of motion. Whereas relativity redefines the
concepts of space and time in terms of the observer, quantum mechanics
denies an aspect of reality to system properties (such as position and
momentum) until they are measured.  This apparent creation of reality
upon measurement is so profound a change that it has engendered an
uneasiness defying formal statement, not to mention a solution.  The
difficulties are often referred to as ``the measurement problem.''
Carried to its logical extreme, the problem is that, if quantum
mechanics were the ultimate theory, it could deny any reality to the
measurement results themselves unless they were observed by yet
another system, ad infinitum. Even the pioneers of quantum mechanics
had great difficulty conceiving of it as a fundamental theory without
relying on the existence of a classical world in which it is embedded
(Landau and Lifshitz 1965).

Quantum mechanics challenges us on another front as well.  From our
intuitive understanding of Bayes' theorem for conditional probability,
we constantly infer the behavior of systems that are observed
incompletely. Quantum mechanics, although probabilistic, violates
Bayes' theorem and thereby our intuition. Yet the very basis for our
concepts of space and time and for our intuitive Bayesian view comes
from observing the natural world. How come the world appears to be so
classical when the fundamental theory describing it is manifestly not
so? This is the problem of the quantum-to-classical transition treated
in this article.

One of the reasons the quantum-to-classical transition took so long to
come under serious investigation may be that it was confused with the
measurement problem. In fact, the problem of assigning intrinsic
reality to properties of individual quantum systems gave rise to a
purely statistical interpretation of quantum mechanics. In this view,
quantum laws apply only to ensembles of identically prepared systems.

The quantum-to-classical transition may also have been ignored in the
early days because regular, rather than chaotic, systems were the
subject of interest. In the former systems, individual trajectories
carry little information, and quantization is straightforward. Even
though Henri Poincar\'e (1992) had understood the key aspects of chaos
and Albert Einstein (1917) had realized its consequences for the
Bohr-Sommerfeld quantization schemes, which were popular at that time,
this subject was never in the spotlight, and interest in it was not
sustained until fairly recently.

As experimental technology progressed to the point at which single
quanta could be measured with precision, the fa\c{c}ade of ensemble
statistics could no longer hide the reality of the counterclassical
nature of quantum mechanics. In particular, a vast array of quantum
features, such as interference, came to be seen as everyday
occurrences in these experiments.

Many interpretations of quantum mechanics developed.  Some appealed to
an anthropic principle, according to which life evolved to interpret
the world classically, others imagined a manifold of universes, and
yet others looked for a set of histories that were consistent enough
for classical reasoning to proceed (Omn\`es 1994, Zurek
2002). However, by themselves, these approaches do not offer a
dynamical explanation for the suppression of interference in the
classical world. The key realization that led to a partial
understanding of the classical limit was that weak interactions of a
system with its environment are universal (Landau and Lifshitz 1980)
and effectively suppress the nonclassical terms in the quantum
evolution (Zurek 1991). The folklore developed that this was the the
only effect of a sufficiently weak interaction in almost any
system. In fact, Wigner functions (the closest quantum analogues to
classical probability distributions in phase space) did often become
positive, but they failed to become localized along individual
classical trajectories.  In the heyday of ensemble interpretations,
this was not a problem because classical ensembles would have been
represented by exactly such distributions.  When applied to a single
quantum system in a single experiment, however, this delocalized
positive distribution is distinctly dissatisfying.

Furthermore, even when a state is describable by a positive
distribution, it is not obvious that the dynamics can be interpreted
as the dynamics of any classical ensemble without hypothesizing a
multitude of ``hidden'' variables (Schack and Caves 1999). And
finally, the original hope that a weak interaction merely erases
interference turned out to be untenable, at least in some systems
(Habib et al. 2000).

The underlying reason for environmental action to produce a
delocalized probability distribution is that if we take a single
classical system with its initial (or subsequent) positions unknown,
the evolving uncertainty in our state of knowledge is encoded by that
distribution. But in an actual experiment, we do know the position of
the system because we continuously measure it. Without this continuous
(or almost continuous) measurement, we would not have the concept of a
classical trajectory. And without a classical trajectory, such
remarkable signals of chaos as the Lyapunov exponent would be
experimentally immeasurable.  These developments brought us to our
current view that continuous measurements provide the key to
understanding the quantum-to-classical transition.

\vspace{-1.2cm}
\section*{\begin{center} Classical vs Quantum Trajectories \end{center}}
\vspace{-0.4cm}

Let us now turn to some significant details. To describe the motion of
a single classical particle, all we need to do is specify a spatially
dependent, and possibly time-dependent, force that acts on the
particle and substitute it into Newton's equations.  The resulting set
of two coupled differential equations, one for the position $x$ of the
particle and the other for the momentum $p$, predicts the evolution of
the particle's state.  If the force on the particle is denoted by
$F(x,t)$, the equations of motion are
\begin{equation} 
\dot{x} = p/m ,
\end{equation}
and
\begin{equation}
\dot{p} = F(x,t) = - \partial_x V(x,t) ,
\end{equation}
where $V(x)$ is the potential.

\begin{figure}[t]
\centerline{\includegraphics[width=6.4in]{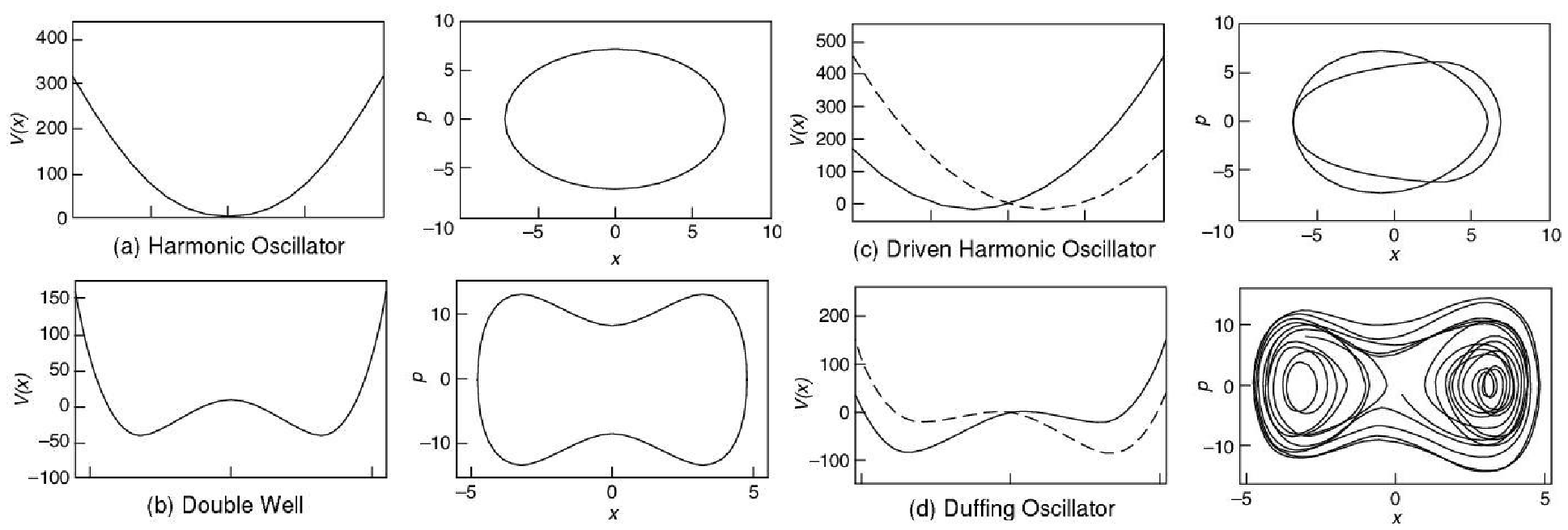}}

\caption{ {\bf Potentials and Phase-Space Trajectories for
Single-Particle Systems } \newline {\small The figure shows four
systems in which a single particle is constrained to move in a
one-dimensional potential. The four systems are (a) a harmonic
oscillator, (b) a double well, (c) a driven harmonic oscillator, and
(d) a driven double well, also known as a Duffing oscillator.  For
each system, the potentials, $V(x)$, are shown next to a typical phase
space trajectory. As the potentials increase in complexity from (a) to
(d), so do the phase-space trajectories.  In (c) and (d), the
potential is time-dependent, oscillating back and forth between the
solid and dashed curves during each period. In (d), the force is
nonlinear, and the trajectory covers increasingly more of the phase
space as time passes.} } 
\label{fig1} 
\end{figure}

To visualize the motion, one can plot the particle's position and
momentum as they change in time. The resulting curve is called a
trajectory in phase space (see Figure~\ref{fig1}).  The axes of phase
space delineate the possible spatial and momentum coordinates that the
single particle can take. A classical particle's state is given at any
time by a point in phase space, and its motion therefore traces out a
curve, or trajectory, in phase space.  By contrast, the state of a
quantum particle is not described by a single point in phase
space. Because of the Heisenberg uncertainty principle, the position
and momentum cannot simultaneously be known with arbitrary precision,
and the state of the system must therefore be described by a kind of
probability density in phase space. This pseudoprobability function is
called the Wigner function and is denoted by $f_{\mbox{\scriptsize
W}}(x,p)$. As expected for a true probability density, the integral of
the Wigner function over position gives the probability density for
$p$, and the integral over $p$ gives the probability density for
$x$. However, because the Wigner function may be negative in places,
we should not try to interpret it too literally. Be that as it may,
when we specify the force on the particle, $F(x,t)$, the evolution of
the Wigner function is given by the quantum Liouville equation, which
is

\begin{equation}
\dot{f}_{\mbox{\scriptsize W}}(x,p) = -\left[ \frac{p}{m} \partial_x
+ F(x,t) \partial_p \right] f_{\mbox{\scriptsize W}}(x,p) +
\sum_{\lambda = 1}^\infty \frac{(\hbar/2i)^{2\lambda}}{(2\lambda + 1)!}   
  \partial_x^{2\lambda+1}
V(x,t) \partial_p^{2\lambda+1} f_{\mbox{\scriptsize W}}(x,p) .
\end{equation}

Clearly, in order for a quantum particle to behave as a classical
particle, we must be able to assign it a position and momentum, even
if only approximately. For example, if the Wigner function stays
localized in phase space throughout its evolution, then the centroid
of the Wigner function~\footnote{The centroid of the Wigner function
is the phase space point defined by the mean values of $x$ and
$p$, that is $(\langle x\rangle, \langle p\rangle)$.} could be
interpreted at each time as the location of the particle in phase
space.

Moreover, the Liouville equation yields the following equations of
motion for the centroid:  
\begin{equation}
\langle \dot{x} \rangle = \langle p \rangle /m ,
\end{equation}
and
\begin{equation}
\langle \dot{p} \rangle = \langle F(x,t) \rangle ,
\end{equation}
where $m$ is the mass of the particle. This result, referred to as
Ehrenfest's theorem, says that the equations of motion for the centroid
formally resemble those for the classical trajectory but differ from
classical dynamics in that the force $F$ has been replaced with the average
value of $F$ over the Wigner function. Suppose again that the Wigner
function is sharply peaked about $\langle x\rangle$ and $\langle
p\rangle$. In that case, we can approximate $\langle F(x)\rangle $ as
a Taylor series expansion about $\langle x\rangle$: 
\begin{equation}
\langle F(x) \rangle = F ( \langle x \rangle ) + \frac{\sigma_x^2}{2}
\partial_x F ( \langle x \rangle ) + \cdots , 
\end{equation}
where $\sigma_x^2$ is the variance of $x$, so that $\sigma_x^2 =
\langle (x - \langle x\rangle)^2\rangle$. If the second and higher
terms in the Taylor expansion are negligible, the equations for the
centroid become 
\begin{equation}
\langle \dot{x} \rangle = \langle p \rangle /m ,
\end{equation}
and
\begin{equation}
\langle \dot{p} \rangle = F(\langle x \rangle,t) .
\end{equation}
And these equations for the centroid are identical to the equation of
motion for the classical particle! If we somehow arrange to start the
system with a sharply localized Wigner function, the motion of the
centroid will start out by being classical, and Equation (6) indicates
precisely how sharply peaked the Wigner function needs to be. 

However, the Wigner function of an unobserved quantum particle rarely
remains localized even if for some reason it starts off that way. In
fact, when an otherwise noninteracting quantum particle is subject to
a nonlinear force, that is, a force with a nonlinear dependence on
$x$, the evolution usually causes the Wigner function to develop a
complex structure and spread out over large areas of phase space. In
the sequence of plots in Figure~\ref{fig2}(a-d), the Wigner function
is shown to spread out in phase space under the influence of a
nonlinear force. Once the Wigner function has spread out in this way,
the evolution of the centroid bears no resemblance to a classical
trajectory.

So, the key issue in understanding the quantum-to-classical transition
is the following: Why should the Wigner function localize and stay
localized thereafter?  As stated in the introduction, this is an
outcome of continuous observation (measurement). We therefore now turn
to the theory of continuous measurements.

\begin{figure}[t]
\centerline{\includegraphics[width=6in]{FullWignerplot.epsc}}
\caption{ {\bf Evolution of the Wigner Function under a Nonlinear
Force} \newline 
{\small These four snapshots show the Wigner function at different
times during a Duffing oscillator simulation. At $t = 0$, the Wigner
function is localized around a single point. As time passes, however,
the Wigner function becomes increasingly delocalized by the
nonlinear potential of the Duffing oscillator. } }
\label{fig2}
\end{figure}  

\vspace{-1.2cm}
\section*{\begin{center} Continuous Measurement \end{center}}
\vspace{-0.4cm}

In simple terms, any process that yields a continuous stream of
information may be termed continuous observation. Because in quantum
mechanics measurement creates an irreducible disturbance on the
observed system and we do not wish to disturb the system unduly, the
desired measurement process must yield a limited amount of information
in a finite time. Simple projective measurements, also known as von
Neumann measurements, introduced in undergraduate quantum mechanics
courses, are not adequate for describing continuous measurements
because they yield complete information instantaneously. The proper
description of measurements that extract information continuously,
however, results from a straightforward generalization of von Neumann
measurements (Davies 1976, Kraus 1983, Carmichael 1993). All we need
to do is let the system interact weakly with another one, such as a
light beam, so that the state of the auxiliary system should gather
very little information about the main one over short periods and
thereby the system of interest should be perturbed only slightly. Only
a very small part of the information gathered by a projective
measurement of the auxiliary system then pertains to the system of
interest, and a continuous limit of this measurement process can then
be taken. By the mid 1990s, this generalization of the standard
measurement theory was already being used to describe continuous
position measurement by laser beams. In our analysis, we use the
methods developed as part of this effort.

A simple, yet sufficiently realistic, analogy to measuring position by
direct observation is measuring the position of a moving mirror by
reflecting a laser beam off the mirror and continuously monitoring the
phase of the reflected light. As the knowledge of the system is
initially imprecise, there is a random component in the measurement
record.  Classically, our knowledge of the system state may be refined
to an arbitrary accuracy over time, and the random component is
thereby reduced. Quantum mechanically, however, the measurement itself
disturbs the system, and our knowledge cannot be improved
arbitrarily. As a result, the measurement record continues to have a
random component.

An equivalent way of understanding this random component is to note
that the measurement process may be characterized by the rate at which
information is obtained. A more powerful measurement is one in which
information is obtained at a faster rate. Because of the Heisenberg
uncertainty relation, if we obtain information about position, we lose
information about momentum. But uncertainty in momentum turns into
uncertainty in position at the very next instant. This random feedback
guarantees that a continuous measurement will cause the system to be
driven by noise: The higher the rate at which information is obtained,
the more the noise. For a position measurement, the rate of
information extraction is usually characterized by a constant, $k$,
that measures how fast the precision in our knowledge of position,
$1/\sigma_x^2$, would increase per unit time in the absence of other
dynamics and the accompanying disturbance. In the laser measurement of
position, $k$ is determined by the power of the laser. The more
powerful the laser, the stronger the measurement, and the more noise
introduced by the photon collisions.

Now we are in a position to see how and under what circumstances
continuous measurement transforms quantum into classical dynamics,
resulting in the quantum-to-classical transition. We can include the
effects of the observation on the motion of the particle by writing
down a stochastic Liouville equation, that is, a Liouville equation
with a random component. This equation is given in the appendix,
Conditions for Approximate Classical Motion under Continuous
Measurement. The resulting equations of motion for the centroid of the
Wigner function are
\begin{equation}
\langle \dot{x} \rangle = \langle p \rangle /m  + \sqrt{8k} \sigma_x^2
\xi(t) , 
\end{equation}
and
\begin{equation}
\langle \dot{p} \rangle = \langle F(x,t) \rangle + \sqrt{8k} C_{xp}
\xi(t) , 
\end{equation}
where $C_{xp} = \langle xp + px\rangle/2 - \langle x\rangle \langle
p\rangle $ is the covariance of $x$ and $p$, and $\xi(t)$ is a
Gaussian white noise. \footnote{White noise is random noise that has
constant energy per unit bandwidth at  every frequency. In reality,
the actual recording of the measurement always occurs at a finite
rate. So, effectively, the white noise gets filtered through a
low-pass filter, which cuts out high frequencies.}  

We have now reached the crux of the quantum-to-classical
transition. To keep the Wigner function well localized, a strong
measurement, or a large $k$, is needed. But Equations (9) and (10)
show that a strong measurement introduces a lot of noise.  In
classical mechanics, however, we deal with systems in which the amount
of noise, if any, is imperceptible compared with the scale of the
distances traveled by the particle. We must therefore determine the
circumstances under which continuous measurement will maintain
sufficient localization for the classical equations to be
approximately valid without introducing a level of noise that would
affect this scale of everyday physics.

With analytical tools alone, this problem cannot be solved. However,
one can take a semianalytical approach by accepting two important
results that come from numerical simulations: (1) Any Wigner function
localizes under a sufficiently strong measurement, and (2) under such
a measurement, once the Wigner function becomes localized, it is
approximately described by a narrow Gaussian at all later
times. Therefore, we assume a Gaussian form for the Wigner function,
write the equations determining how the variances and covariances
change with time, and solve those equations to find their values in a
steady state. Having all these ingredients, we can then find the
conditions under which the noise terms are small and the system
remains well localized (see the Appendix). Our central conclusion is
that a quantum system will behave almost classically for an
ever-increasing range of measurement strengths when the action of the
system is large compared with the reduced Planck constant $\hbar$.

This concept may be understood heuristically in the following way:
Because of the uncertainty principle, the effective area where the
localized Gaussian Wigner function is nonzero can never be less than
$h$. If this limiting area is so large compared with the scale of the
problem that it cannot be considered localized, we certainly do not
expect classical behavior. Conversely, as long as the measurement
extracts information at a sufficiently low rate to avoid squeezing the
Wigner function to a smaller scale than the limiting one, the quantum
noise remains on the scale of the variances themselves. As a result,
the system behaves almost classically.

There are systems, however, whose phase space is sufficiently small
for quantum effects to be manifest or even dominant. This is true, for
example, of isolated spin systems with small total angular
momenta. Even when they are observed and interacting with the
environment, these spin systems are expected to be indescribable by
the classical laws of motion. A spin coupled to other degrees of
freedom such as position is a more interesting case, especially when
the position of the system would have followed a classical trajectory
in the absence of that interaction. To what extent, if at all, that
coupling stops position from following a classical trajectory is the
subject of ongoing research (Ghose et al. 2003).

\vspace{-1.2cm}
\section*{\begin{center} Chaos in a Quantum System under Continuous
Observation \end{center}} 
\vspace{-0.4cm}

As an illustration of these general ideas, we consider the Duffing
oscillator, a single particle sitting in a double-well potential and
driven sinusoidally -- see Figure~\ref{fig1}(d).  We chose this nonlinear
system because it has been studied in depth and it allows us to choose
parameters that produce chaotic behavior over most of the system's
phase space.  Our test will indicate whether chaotic classical motion
is a good approximate description of this quantum system when it is
under continuous observation. To diagnose the presence of chaos, we
calculate the maximal Lyapunov exponent, the most rigorous measure of
chaotic behavior~\footnote{The maximal Lyapunov exponent is one of a
number of coefficients that describe the rates at which nearby
trajectories in phase space converge or diverge.}, and compare our
calculated value for the quantum system with the classical value.

The Hamiltonian for the particle in the double-well potential is
\begin{equation}
  H = \frac{p^2}{2m} + B x^4 - A x^2 + \Lambda x \cos(\omega t) ,
\end{equation}
where $m, A, B, \Lambda$, and $\omega$ are parameters that determine
the size of the particle and the spatial extent of the phase
space. The action should be large enough so that the particle can
behave almost classically, yet small enough to illustrate how tiny it
needs to be before quantum effects on the particle become
dominant. Bearing this requirement in mind, we choose a mass $m = 1$
picogram, a spring constant $A = 0.99$ piconewton per meter, a
nonlinearity $A/B = 0.02$ square micrometer, a peak driving force of
$\Lambda = 0.03$ attonewton, and a driving frequency $\omega = 60$ rad
per second.  Because of the weakness of the nonlinearity, the distance
between the two minima of the double well is only about 206
nanometers, and the height of the potential is only 33 nano-eV.  The
frequency of the driving force is 10 hertz. For these values, a
measurement strength $k$ of 93 per square picometer per second, which
corresponds to a laser power of about 0.24 microwatt, is adequate to
keep the motion classical, or the Wigner function well-localized.

\begin{figure}[t]
\centerline{\includegraphics[width=6.4in]{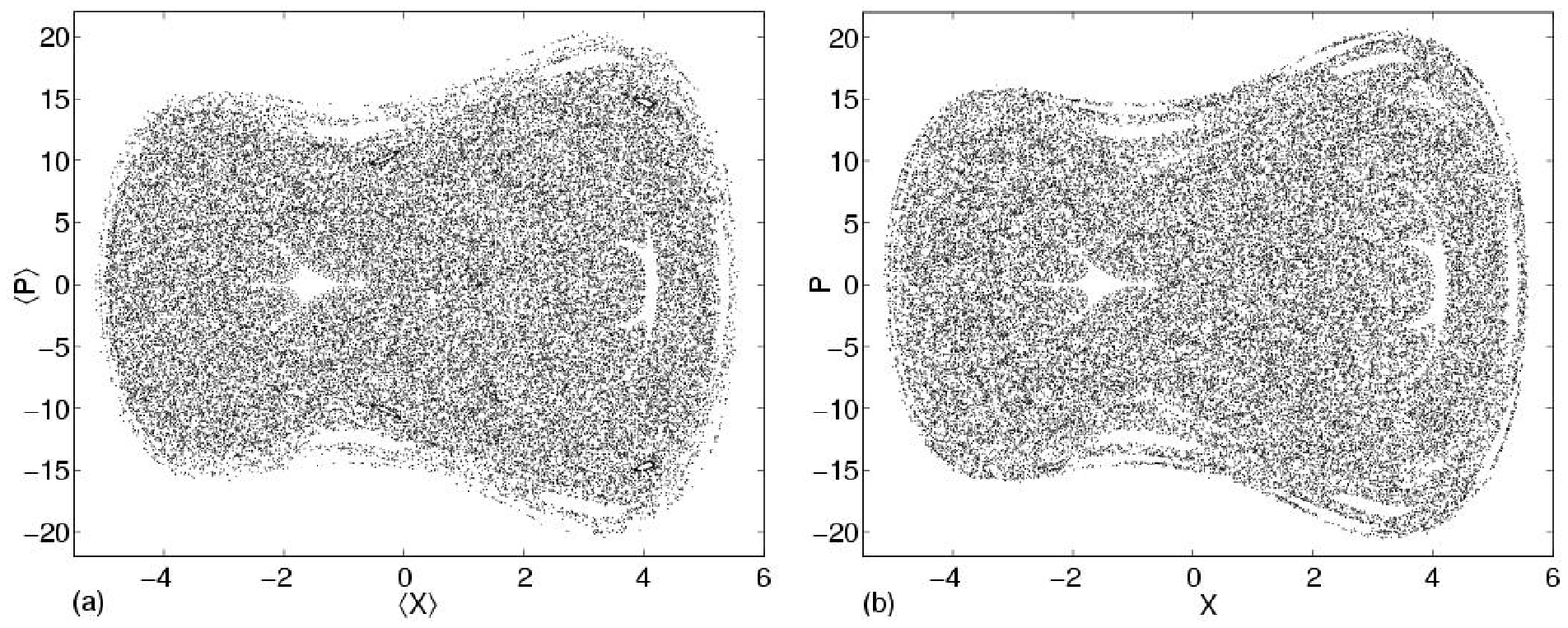}}
\caption{ {\bf Stroboscopic Maps for the Quantum and Classical Duffing
Oscillators} \newline
{\small The results of the Duffing oscillator simulations are plotted
as stroboscopic maps. (a) The map for the continuously observed
quantum Duffing oscillator displays the centroids of the Wigner
function at time intervals separated by the period of the driving
force. This map is a pastiche from several different runs with
different initial conditions, for a total duration of 39,000 periods
of the temporal drive. (b) The map for the classical Duffing
oscillator driven with a small amount of noise displays the calculated
locations of particles in phase space at time intervals separated by
the period of the driving force. The two maps are very similar. The
quantum system under continuous measurement exhibits qualitatively the
same chaotic behavior as the classical system driven with a small
amount of noise. In these figures, $\Delta X = 33$ nm, and $\Delta P =
324$ pg nm/s. } } 
\label{fig3}
\end{figure}

To study the system numerically, we allow the particle's Wigner
function to evolve according to the stochastic Liouville equation for
approximately 50 periods of the driving force and then check that it
remains well localized in the potential. We find, indeed, that the
width of the Wigner function in position (given by the square root of
the position variance $\sigma_x^2$) is always less than 2
nanometers. Thus the position of the particle is always well resolved
by the measurement as the system evolves. In addition, an inspection
of the centroid's trajectory shows that the noise is negligible. In
order to verify that the motion is, in fact, that of a classical
Duffing oscillator, we perform two tests.  The first is to plot a
stroboscopic map showing the particle's motion in phase space and then
compare that map with the corresponding one of the classical Duffing
oscillator driven by a small amount of noise. The observed quantum map
and the classical map are displayed in Figure~\ref{fig3}.

The two stroboscopic maps are very similar and show qualitatively that
the quantum dynamics under continuous measurement exhibits chaotic
behavior analogous to classical chaos. To verify this finding
quantitatively, we conduct a second test and calculate the Lyapunov
exponent for both systems. As we already mentioned, trajectories that
are initially separated by a very small phase-space distance,
$\Delta(0)$, diverge exponentially as a function of time in chaotic
systems. The Lyapunov exponent $\lambda$, which determines the rate of
this exponential increase, is defined to be
\begin{equation}
\lambda = \lim_{t\rightarrow\infty} \lim_{\Delta(0)\rightarrow 0}
\frac{\ln \Delta(t)}{t} . 
\end{equation}

\begin{figure}[t]
\centerline{\includegraphics[width=5in]{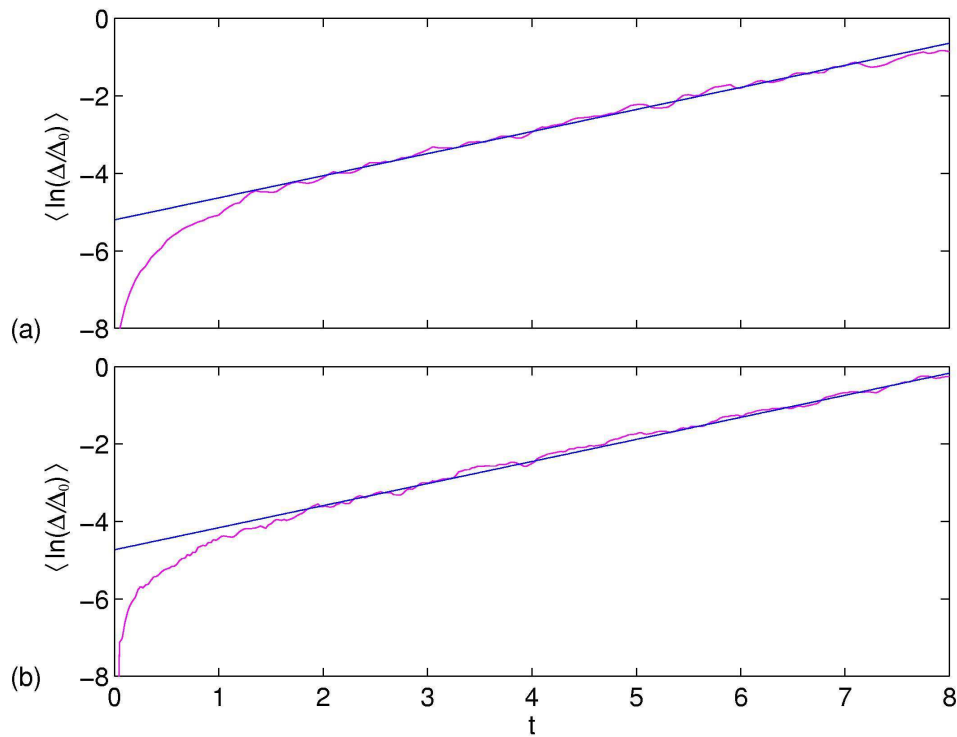}}
\caption{ {\bf Lyapunov Exponents for the Quantum and Classical
Duffing Oscillators} \newline 
{\small In order to calculate the Lyapunov exponents, $\lambda$, for
(a) a classical Duffing oscillator driven with a small amount of
noise, and (b) a continuously observed quantum Duffing oscillator, we
plot against time the logarithm of the average separation of
trajectories that begin very close together. The parameters defining
the oscillator -- the continuous-measurement strength in the quantum
system and the noise in the classical system -- have been detailed
earlier. The slope of the line drawn through the curves gives the
Lyapunov exponent, which in both cases is $\lambda = 0.57(2)$.
Also in both cases, $\Delta_0 = 33$ nm.} }
\label{fig4}
\end{figure}

To calculate this exponent, we first choose a single fiducial
trajectory in which the centroid of the Wigner function starts at the
phase-space point given by $\langle x \rangle = Ð 98$ nanometers and
$\langle p\rangle = 2.6$ picograms micrometers per second (pg
$\mu$m/s).  At 17 intervals along this trajectory, each separated by
approximately 20 periods of the driving force, we obtain neighboring
trajectories by varying the noise realization.  We calculate how these
trajectories diverge from the initial trajectory and average the
difference over the 17 sample trajectories. We then carry out this
procedure for 10 fiducial trajectories, all starting at the same
initial point but differing because of different noise
realizations. Plotting the logarithm of this average divergence as a
function of time results in a line whose slope is the Lyapunov
exponent. In Figure~\ref{fig4}, we plot the logarithm of the average
divergence for both the observed quantum system and the classical
system driven with a small amount of noise. The slope of the lines
drawn through the curves gives the Lyapunov exponent, which in both
cases is 0.57(2) per second. To show that the noise has a negligible
effect on the dynamics, we also calculate the Lyapunov exponent for
the classical system with no noise, using trajectories starting in a
small region around the point given by $x = -98$ nanometers and $p =
2.6$ pg $\mu$m/s. Those trajectories give a Lyapunov exponent of
0.56(1) per second, which is in agreement with the previous value.

Now we elaborate on the problem hinted at in the introduction. If
observation realizes the classical world, do trees in remote forests
fall quantum mechanically?  Of course, the tongue-in-cheek answer is,
``who knows?'' At a deeper level, however, we note that even in a
remote forest, trees continue to interact with the environment, and
through this interaction, the components of the environment (reflected
light, air molecules, and so on) acquire information about the
system. According to unitarity, an important property of quantum
mechanics, information can never be destroyed. The information that
flowed into the environment must either return to its origin or stay
somewhere in the environment -- the decaying sound of the falling tree
must yet record its presence faithfully, albeit perhaps only in a
shaken leaf. And herein lies the key to understanding the unobserved:
If a sufficiently motivated observer were to coax the information out
of the environment, that action would become an act of continuous
measurement of the current happenings even though actually performed
in the future.  But since the current state of affairs cannot be
influenced by what anyone does in the future, the behavior of the
system at present cannot contradict anything that such a classical
record could possibly postdict.

If the motion is not observed, no one knows which of the possible
paths the object took, but the rest of the universe does record the
path, which could, therefore, be considered as classical as any
(Gell-Mann and Hartle 1993). All that happens when there is no
observer is that our knowledge of the motion of the object is the
result of averaging over all the possible trajectories. In that case,
we are forced to describe the state of the system as being given by a
probability distribution in phase space since we no longer know
exactly where the system is as it evolves. This observation is,
however, just as true for a (noisy) classical system as it is for a
quantum system.

\vspace{-1.2cm}
\section*{\begin{center} The Connection to the Theory of Decoherence
\end{center}} 
\vspace{-0.4cm}

We can now explain how the analysis presented here relates to a
standard approach to the quantum-to-classical transition often
referred to as decoherence. The procedure employed in decoherence
theory is to examine the behavior of the quantum system coupled to the
environment by averaging over everything that happens to the
environment.  This procedure is equivalent to averaging over all the
possible trajectories that the particle might have taken, as explained
above. Thus, decoherence gives the evolution of the probability density
of the system when no one knows the actual trajectory.  The relevant
theoretical tools for understanding this process were first developed
and applied in the 1950s and 1960s (Redfield 1957, Feynman and Vernon
1963), but more recent work (Hepp 1972, Zurek 1981, 1982, Caldeira and
Leggett 1981, 1983a, 1983b, Joos and Zeh 1985) was targeted at
condensed-matter systems and a broader understanding of quantum
measurement and quantum-classical correspondence. It was found that
averaging over the environment or over the equivalent, unobserved,
noisy classical system gives the same evolution (Habib et
al. 1998). In this classical counterpart, different realizations of
noise give rise to slightly different trajectories, and in a chaotic
system, these trajectories diverge exponentially fast. As a result,
probability distributions obtained by averaging over the noise tend to
spread out very fast, and our knowledge of the system state is
correspondingly reduced. In other words, discarding the information
that is contained in the environment or, equivalently, the measurement
record, as averaging over these data implies, leads to a rapid loss of
information about the system. This increasing loss of information,
characterized by a quantity called entropy, can then be used to study
the phenomenon of chaos with varying degrees of rigor.

Averaging over the environment to produce classical probability
distributions was, however, not completely satisfactory. Not only does
this averaging procedure not allow us to calculate trajectory-based
quantities, but it also restricts our predictions to those derivable
by knowing only the probability densities at various times. But
classical physics is much more powerful than that -- it can predict
the outcome of many ``if ... then'' scenarios. If I randomly throw a
ball in some direction, the probability of it landing in any direction
around me is the same, but if you see the ball north of me, you can
predict with pretty good certainty that it won't land south of me. In
the classical world, such correlations are numerous and varied, and
the measurement approach we have taken here completes our
understanding of the quantum-to-classical transition by treating all
correlations on an equal footing. It is easy to see, however, that if
the continuous measurement approach has to get all the correlations
right, it must per force get the decoherence of probability densities
right!

The realization that continuous measurement was the key to
understanding the quantum-to-classical transition has emerged only in
the last decade. First introduced in a paper by Spiller and Ralph
(1994), this idea was then mentioned again by Schlautmann and Graham
(1995). Subsequently, the idea was developed in a collection of papers
(Schack et al. 1995, Brun et al. 1996, Percival and Strunz 1998,
Strunz and Percival 1998). However, the scientific community was slow
to pick up on this work, possibly because the authors used a
stochastic model referred to as quantum state diffusion, which may
have obscured somewhat the measurement interpretation. In 2000, we
published the results presented in this article, namely, analytic
inequalities that determine when classical motion will be achieved for
a general single-particle system, and showed that the correct Lyapunov
exponent emerges (Bhattacharya et al. 2000). For this purpose, we used
continuous position measurement, which is ever present in the everyday
world and therefore the most natural one to consider. This
accumulation of work now provides strong evidence that continuous
observation supplies a natural and satisfactory explanation for the
emergence of classical motion, including classical chaos, from quantum
mechanics. In addition, such an analysis also makes clear that the
specific measurement model is not important. Any environmental
interaction that provides sufficient information about the location of
the system in phase space will induce the transition in macroscopic
systems. Scott and Milburn (2001) have analyzed the case of continuous
joint measurement of position and momentum and of momentum alone, and
they verified that classical dynamics emerges in the same way as
described in Bhattacharya et al. (2000).

\appendix

\section*{{\large Appendix: Conditions for Approximate Classical
Motion}}

The evolution of the Wigner function $f_W$ for a single particle
subjected to a continuous measurement of position is given by the
stochastic Liouville equation: 
\begin{eqnarray}
f_{\mbox{\scriptsize W}}(x,p,t+dt) \!\!\! & = & \!\!\! \left[ 1 - dt
\! \left[ \frac{p}{m} \partial_x  + F(x,t) \partial_p \right] 
+ dt \! \sum_{\lambda = 1}^\infty \! \frac{\left(
\hbar/2i\right)^{2\lambda}}{(2\lambda + 1)!}  \partial_x^{2\lambda+1}
V(x,t) \partial_p^{2\lambda+1} \right] f_{\mbox{\scriptsize W}}(x,p,t)
\nonumber \\
& + & \sqrt{8k} \xi(t) dt (x - \langle x \rangle )
f_{\mbox{\scriptsize W}}(x,p,t) , 
\end{eqnarray}
where $F$ is the force on the particle, $\xi(t)$ is a Gaussian white
noise, and $k$ is a constant  characterizing the rate of information
extraction. Making a Gaussian approximation for the Wigner function,
which according to numerical studies is a good approximation when
localization is maintained by the measurement, the equations of motion
for the variances of $x$ and $p$, $\sigma_x^2$ and $\sigma_p^2$, are
\begin{eqnarray}
\dot{\sigma}_x^2  & = & \frac{2}{m} C_{xp} - 8k \sigma_x^4 \, ,
\;\;\;\;\;\;\;\;\;\;\;   
\mbox{where} \;\; C_{xp} = \langle xp + px\rangle/2 - \langle x\rangle
\langle p\rangle  , \label{c1} \\  
\dot{\sigma}_p^2  & = & 2\hbar^2 k - 8k C_{xp}^2 + 2\partial_x F
C_{xp} \, , \label{c2} \\   
\dot{C}_{xp} & = &  \frac{1}{m} \sigma_p^2 - 8k \sigma_x^2 C_{xp} +
\partial_x F \sigma_x^2 \label{c3}\, , 
\end{eqnarray}
the noise has negligible effect in these equations when the Wigner
function stays Gaussian. First, we solve these equations for the
steady state and then impose on this solution the  conditions required
for classical dynamics to result. In order for the Wigner function to
remain sufficiently localized, the measurement strength $k$ must stop
the spread of the wave function at the unstable points, $\partial_x F
> 0$:\footnote{If the nonlinearity is large on the quantum scale
$\hbar |(\partial^2_x F)/F| \ge 4\sqrt{m|\partial_x F|}$, then $8k$
needs to be much larger than $(\partial^2_x F)^2\hbar/4mF^2$
irrespective of the sign of $\partial_x F$. This observation does not
change the argument in the body of the paper.} 
\begin{equation} 
 8k \gg \left| \frac{\partial_x^2 F}{F} \right| \sqrt{\frac{\left|
\partial_x F \right|}{2m}} . 
\end{equation}
If noise is to bring about only a negligible perturbation to the
classical dynamics, it is sufficient that, at a typical point on the
trajectory, the measurement satisfy 
\begin{equation}
\frac{2 \left| \partial_x F \right|}{s} \ll \hbar k \ll \frac{\left|
\partial_x F \right| s}{4} , 
\end{equation}
where $s$ is the typical value of the systemÕs action~\footnote{We are
assuming that both $[mF^2/(\partial_x F)^2] |F/p| $ and $E |p/4F|$
evaluated at a typical point of the trajectory are comparable to the
action of the system, and we define that action to be $\hbar s$.} in
units of $\hbar$. Obviously, as $s$ becomes much larger than
$2\sqrt{2}$ this relationship is satisfied for an ever-larger range of
$k$. At the spot where this range is sufficiently large, we obtain the
classical limit.

\end{document}